\documentclass [12pt]{article}
\usepackage {graphicx}
\usepackage {amssymb}
\usepackage {amsmath}
\usepackage {longtable}

\title{Modified Planck units}
\author{$^{1}$\textbf{Yu.L.Bolotin}, $^{2,3}$\textbf{V.V.Yanovsky} }

\begin{document}

 \maketitle

$^{1}$\textit{A.I.Akhiezer Institute for Theoretical Physics, National Science Center "Kharkov Institute of Physics and Technology",
NAS Ukraine, Akademicheskaya Str. 1, 61108 Kharkov, Ukraine}

$^{2}$ \textit{Institute for Single Crystals, NAS Ukraine, Nauky Ave. 60, Kharkov 31001, Ukraine}

$^{3}$\textit{ V. N. Karazin Kharkiv National University, 4 Svobody Sq., Kharkiv, 61022, Ukraine}

\bigskip

\begin{abstract}Planck units are natural physical scales of mass, length and time, built with the help of the fundamental constants $\hbar, c, G$. The functional role of the constants used for the construction of Planck units is different. If the first two of them represent the limits of the action and the speed of light and underlie quantum mechanics and special relativity, the Newton's constant $G$  "only" fixes the absolute value of the gravitational forces. It seems natural to make a set of fundamental constants more consistent and more effective if used to build Planck units only limit values.  To this end, in addition to the limit values $\hbar $  and $c$ we introduce an additional limit value - a maximum power in nature. On the basis of these values, a modification of the Planck unit system is proposed. The proposed modification leaves unchanged the numerical values of Planck units, however, opens up exciting new possibilities for interpreting the known results and for obtaining  new ones.
\end{abstract}

\section{Introduction}

Introduction of the Planck system of units  \cite{s1} prior to creation of quantum mechanics as well as of special and general relativity theory, was a significant event in physics.  The   system of "natural measurement units", as Planck called it,  put forward the questions which were significantly ahead of the state of physics at that time.  To answer these questions, there had to be achieved a considerable progress in the development of physics. Now, more than a century later, we are only groping for approaches to the raised problems, and the final understanding seems to be achieved solely after creation of quantum theory of gravitation. The systems of units based on different fundamental constants not only reflect the history of the development of physics, but also permit to estimate the prospects  of this development. Thereat,  the Planck units still  give an opportunity to look into the future.

Various problems bound up with the Planck system of units are considered in a great number of papers  (see e.g. \cite{s2,s3,s4,s5,s6}) in the field of elementary particle physics and cosmology. Among such works one should mention  those devoted to attempts to build alternative sets of fundamental scales which differ  both in the choice of constants and in their number \cite{s7,s8,s9,s10}.

In the present paper, a new modification of the Planck system of units is proposed. While considering the fundamental constants which belong to this system of units it should be emphasized that the functional role of the constants  $\hbar, c$ and $G$ is different. The constants $\hbar$ and $c$  are limit values and underlie quantum mechanics $(\hbar)$ and special relativity $(c)$, whereas the Newton constant $G$ "merely" fixes the interaction strength. It seems natural to make the set of fundamental constants more consistent and more effective for subsequent analysis of physics on the Planck scale, if to replace the gravitation constant  $G$ by a new limit value associated with general relativity. In this case the Planck units will be expressed only in terms of the limit values  which underlie quantum mechanics $(\hbar)$, Special Relativity $(c)$ and General Relativity (maximum force or maximum power). At such transition   the numerical values of the Planck units remain unchanged, however there appear new interesting  opportunities  for interpretation of well-known results and for obtaining  new ones.

In this paper we discuss advantages of the proposed modification and give a number of examples to illustrate its effectiveness

\section{Modification of Planck  system of units}

Dimensional analysis is a powerful method which makes it possible to obtain results  (both qualitative and quantitative) on the base of general knowledge of the  phenomenon under consideration  Dimensional analysis (along with symmetry considerations) is especially significant  for construction of initial approaches to description of those systems for which any theory is absent at present. As is well-known, many astonishing results have been achieved owing to  dimensional considerations which at first sight seem to be quite simple. There are even such results are still not obtained by other, more rigorous way. A classic illustration of such a situation is quantum gravity. The latter has practically  become a synonym of Planck-scale physics which description to a considerable extent reduces to endless shuffle of fundamental constants. However, dimensional analysis is not all-powerful, and the results obtained with its help should be interpreted carefully.

Especially significant role in clarification/understanding of the foundations of the future theory Planck-scale processes belongs to the same name units. The Planck units are fundamental physical scales of mass, length and time built by means of the fundamental constants $\hbar ,c,G$:
	\[\begin{array}{l}
{m_{Pl}} = \sqrt {\frac{{\hbar c}}{G}}  \simeq 2.18 \times {10^{ - 8}}kg,\\
{l_{Pl}} = \sqrt {\frac{{\hbar G}}{{{c^3}}}}  \simeq 1.6 \times {10^{ - 35}}m,\\
{t_{Pl}} = \sqrt {\frac{{\hbar G}}{{{c^5}}}}  \simeq 5.39 \times {10^{ - 44}}\sec
\end{array}\]
When presenting new universal constants in his report for Prussian Academy of Sciences in Berlin  (1899), Planck noted: "These units preserve their natural value till the law of gravity as well as both laws of thermodynamics hold true,\footnote{besides  $\hbar ,c,G$, Planck considered  the Boltzmann constant to be a universal one, too} and till the speed of light propagation  in the vacuum  remains invariable. Therefore, though measured by a variety of intellectuals  and using  different methods, they will always have the same value".

It should be noted that the choice of  $\hbar ,c,G$ in the capacity of initial constants necessary for construction of  fundamental scales, is not unique. As an initial material, there may be used various combinations of these constants. One can give many examples of the cases when the use of one or another combination of fundamental constants essentially simplifies the  theory. One of typical examples is  the fine-structure constant  $\alpha  = {e^2}/\hbar c$. It is difficult to imagine quantum  electrodynamics in which this combination is not used.

Our choice of the starting constants is based on the following considerations. It should  be reminded that the statement concerning the existence of limit values may be a foundation  of physical  axiomatics.  As is well-known, quantum mechanics can be constructed proceeding from the existence of the minimum quantum of action $\hbar $, whereas Special Relativity  implies the existence of the maximum speed $c$. Not long ago it has become clear that an analogous approach may be also realized in General Relativity  if to postulate the existence of a maximum force \cite{s11}. The proof  based on the Jacobson's argumentation \cite{s12} shows that the Einstein equations can be derived from thermodynamic identities on the Rindler horizon.  Gibbons \cite{s13} postulated the existence of limit  (maximum) force  in the  principle: «I suggest that classical General Relativity in four spacetime dimensions incorporates a Principal of Maximal Tension and give arguments to show that the value of the maximal tension is»
\begin{equation}\label{e1}
  {F_{\max }} = \frac{{{c^4}}}{{4G}} \approx 3.25 \times {10^{43}}\;N
\end{equation}
The limit does not depend on the nature of forces and holds for gravitational, electromagnetic, nuclear and any other forces. Naturally, this value is independent of the choice of  reference system  (inertial or non-inertial), and may be interpreted for our World as an additional  (on a par with   $\hbar $ and $c$ ) fundamental limit constant.

Alternatively, it is possible to use as a basic principle equivalent statement as a basic principle:
there is a maximum power,
\begin{equation}\label{e2}
  {P_{\max }} = \eta  = \frac{{{c^5}}}{{4G}} \approx 9.07 \times {10^{51}}W
\end{equation}
Both values are the components of the 4-vector  ${F^\lambda } = \frac{{d{p^\lambda }}}{{dt}}$. The limit power admits a trivial physical interpretation. Let us consider the power released in the process of  "annihilation" of a black hole with the mass $M$. The minimum duration of this process is the  time  $t = 2{R_g}/c = 4MG/{c^3}$, where ${R_g} = 2MG/{c^2}$ is the gravitational radius.     The power released in such a process is expressed as
\begin{equation}\label{e3}
  P = \frac{{M{c^2}}}{{4MG/{c^3}}} = \frac{{{c^5}}}{{4G}} = \eta
\end{equation}
The multiplier  $1/4$ does not play a principal role. Therefore, further,  where it does not lead to confusion,  we will omit numerical multipliers of the order of unity  assuming that the approximate equality $A \approx B$ corresponds to the relation $\lg A \approx \log B$.

The mechanism of the occurrence of the limit force (power) is extremely simple \cite{s13}. Let us consider two bodies with the masses  ${M_1}$ and ${M_2}$, the distance between  which being $R$. According to the Newton's theory, the force  acting between them  are expressed as
\begin{equation}\label{e4}
  F = G\frac{{{M_1}{M_2}}}{{{R^2}}} = \left( {\frac{{G{M_1}}}{{{c^2}R}}} \right)\left( {\frac{{G{M_2}}}{{{c^2}R}}} \right)\frac{{{c^4}}}{G}
\end{equation}
Since
 \[{M_1}{M_2} \le \frac{1}{4}{({M_1} + {M_2})^2} ,\]
then
\begin{equation}\label{e5}
  F \le {\left[ {\frac{{\left( {{M_1} + {M_2}} \right)G}}{{{c^2}R}}} \right]^2}\frac{{{c^4}}}{{4G}}
\end{equation}
Approaching of the bodies is limited by the condition  $R > {R_g}$ $\left(R_g =\frac{2MG}{c^2} \right)$ which prevents  the formation of a black hole with the mass  ${M_1} + {M_2}$. Therefore,
\begin{equation}\label{e6}
  F \le \frac{{{c^4}}}{{4G}}
\end{equation}
Surfaces on which implemented the maximum force (maximum pulse flow), or the maximum power (maximum energy flow) are horizons. Any attempt to exceed  the force limit leads to the appearance of  horizon. In its turn, the latter does not permit to exceed the limit.

Here is another example to  clarify the mechanism of the occurrence of the maximum force. In the Newtonian mechanics $F = dp/dt$, therefore
\begin{equation}\label{e7}
  {F_{max}} = \frac{{{{(\Delta p)}_{\max }}}}{{{{\left( {\Delta t} \right)}_{\min }}}} \approx \frac{{mc}}{{{t_{Pl}}}} = \frac{{m{c^2}}}{{{l_{Pl}}}}
\end{equation}
At first sight one  may expect that unlimited growth of the mass will give rise  to an arbitrarily great  force.   However, this is not so, and the limitation is bound up with the appearance of  horizon  at the increase of the mass on a fixed scale of length $\left( {{l_{Pl}}} \right)$. Indeed, when omitting the numerical multipliers ${\rm O}(1)$ we find the mass with the gravitational radius equal to the Planck length:
\begin{equation}\label{e8}
  m \approx \frac{{{l_{Pl}}{c^2}}}{G} = \sqrt {\frac{{\hbar G}}{{{c^3}}}} \frac{{{c^2}}}{G} = \sqrt {\frac{{\hbar c}}{G}}  = {m_{Pl}}
\end{equation}
Consequently, the maximum mass which can be used in (\ref{e7}) for preventing the appearance of the horizon is the Planck mass, so
\begin{equation}\label{e9}
  {F_{\max }} \approx \frac{{{m_{Pl}}{c^2}}}{{{l_{Pl}}}} = \frac{{{c^4}}}{G}
\end{equation}
It is surprising that the result (\ref{e9}) can be obtained in the form of the combination of the Planck units with the dimension of  force:
\begin{equation}\label{e10}
  {F_{Pl}} = {m_{Pl}}\frac{{{l_{Pl}}}}{{t_{Pl}^2}} = \sqrt {\frac{{\hbar c}}{G}} \sqrt {\frac{{\hbar G}}{{{c^3}}}} \frac{{{c^5}}}{{\hbar G}} = \frac{{{c^4}}}{G}
\end{equation}
Note that for the Planck mass the gravitational radius coincides with the Compton wavelength.
It should be emphasized that all our statements concern solely $D = N + 1 = 4$. It is only in the space $D = 4$  that the Planck force is independent of $\hbar $:
\begin{equation}\label{e11}
  {F_{Pl(D)}} = \frac{{{M_{Pl(D)}}{L_{Pl(D)}}}}{{T_{Pl(D)}^2}} = G_D^{ - \frac{2}{{D - 2}}}{\hbar ^{D - 4}}{c^{\frac{{D + 4}}{{D - 2}}}}
\end{equation}
More strictly the expression for the limit force (limit power) can be obtained in the frames of GR \cite{s11}.

The force on a test mass $m$  at a radial distance $d$  from a Schwarzschild black hole is
\begin{equation}\label{e12}
  F = G\frac{{Mm}}{{{d^2}\sqrt {1 - \frac{{2GM}}{{{c^2}d}}} }}
\end{equation}
At first sight it seems that this relation contains a singularity and makes it possible to achieve an arbitrarily large value of force. However, if to discard non-physical point-like masses, it turns out that the limit value of force exists. Indeed, any mass cannot reduce its size down to the gravitational radius.   Otherwise it will be transformed into a black hole, and it will be unobservable. The minimum distance is the sum of the gravitational radii of these masses $d \ge {d_{\min }} = \frac{{2mG}}{{{c^2}}} + \frac{{2MG}}{{{c^2}}} = \frac{{2G\left( {m + M} \right)}}{{{c^2}}}$. Having  substituted the minimum distance into  (\ref{e12}) we obtain
\begin{equation}\label{e13}
  F = \frac{{{c^4}}}{{4G}}\frac{{Mm}}{{{{\left( {M + m} \right)}^2}}}\frac{1}{{\sqrt {1 - \frac{M}{{M + m}}} }} \le \frac{{{c^4}}}{{4G}}
\end{equation}
Now consider the use of the limit power  for construction of modified Planck units. To this end we replace the gravitational constant   by the limit power, $G = \frac{{{c^5}}}{\eta }$ . In other words, by using  the set  $\left( {\hbar ,c,\eta } \right)$ instead of $\left( {\hbar ,c,G} \right)$ we get the modified system of Planck units  (which consists only of  the limit values):
\[{m_{Pl}} = \sqrt {\frac{{\hbar \eta }}{{{c^4}}}}  \simeq 2.18 \times {10^{ - 8}}kg,\]
\[{l_{Pl}} = \sqrt {\frac{\hbar }{\eta }} c \simeq 1.6 \times {10^{ - 35}}m,\]
\begin{equation}\label{e14}
  {t_{Pl}} = \sqrt {\hbar /\eta }  \simeq 5.39 \times {10^{ - 44}}\sec
\end{equation}
It is interesting to note that the Planck time is defined only by two of the  introduced fundamental constants. That is, by changing $c$ and $G$ concordantly in such a way that  $\eta$ preserves its value one can preserve the value of the Planck time, herewith changing the values of the Planck mass and  length. Naturally, such a possibility, though less conspicuous, also existed for the original the Planck choice  of fundamental constants.

It should be emphasized again that the necessary condition for the existence of a event horizon  is finiteness of the realized power and  of the speed of light. Thereat, as we have already noted, the magnitude  of the limit value is less significant than the fact of its existence. It is easily seen that
\[\mathop {\lim }\limits_{c \to \infty } {R_g} = \mathop {\lim }\limits_{c \to \infty } \frac{{2mG}}{{{c^2}}} = 0;\]
\begin{equation}\label{e15}
  \mathop {\lim }\limits_{{P_{\max }} \to \infty } {R_g} = \mathop {\lim }\limits_{c \to \infty } \frac{{2m{c^3}}}{{{P_{\max }}}} = 0
\end{equation}
In other words, at  $\eta  \to \infty $ or $c \to \infty $ the concept of gravitational radius and, consequently, the event horizon, becomes meaningless.

While considering the  maximum force (maximum power) as a fundamental constant it is natural to use it instead of the gravitational constant. For instance,  Newton's law of universal gravitation acquires the form
\[F = G\frac{{mM}}{{{R^2}}} = \frac{{{c^4}mM}}{{4{F_{\max }}{R^2}}} = \frac{1}{{{F_{\max }}}}\frac{{m{c^2} \cdot M{c^2}}}{{{R^2}}}\]
Here the relation between the value of gravitational interaction and the maximum force becomes more transparent: a gigantic  maximum force gives rise to  a weak gravitational interaction.  Naturally, if to choose the gravitational constant  $G$ in the capacity of  initial fundamental constant, the converse statement is true, too.

The choice of the maximum power as a new fundamental constant leads to the Planck scales  which preserve their  previous numerical values. However, such a changeover opens up interesting opportunities for interpretation of  estimations made using the modified Planck units, as well as for the obtaining of new results. Below we will give a number of examples.

\section{Space-time foam}

If a space undergoes quantum fluctuations, the latter must manifest themselves as uncertainties in different kinds of measurements \cite{s14,s15,s16}. Among them, measurement of lengths is   significant.  Let  $\delta l$ is  an uncertainty with which the length$l$  can be measured. To solve this problem, Wigner \cite{s17,s18} proposed a gedanken experiment. Let us place a clock   and a mirror at different ends of the segment to be measured. By measuring the time of registration of a  reflected signal one can measure the length of the segment. However, quantum fluctuations will generate the uncertainty $\delta l$ for the distance to be measured. Wigner showed that
\begin{equation}\label{e16}
  \delta {l^2} \ge \frac{{\hbar l}}{{mc}}
\end{equation}
Here  $m$ is the clock mass. As first sight it seems that the influence of quantum fluctuations will  be eliminated if the clock mass tends to infinity. However, the growth of the mass is strictly limited.  It is obvious that  the watch size  $d$ is limited   by the condition of the experiment $\left( {d \le \delta l} \right)$. On the other hand, the size of the clock must exceed its  Schwarzschild radius $d > Gm/{c^2}$ to prevent transformation of the clock into a black hole, since otherwise the indications of the clock will be inaccessible  for observers. From here it follows that
\begin{equation}\label{e17}
  \delta l \ge \frac{{Gm}}{{{c^2}}}
\end{equation}
By combining  (\ref{e14})  and (\ref{e15}) we obtain
\begin{equation}\label{e18}
  \delta l \ge {\left( {ll_{Pl}^2} \right)^{1/3}} = {l_{Pl}}{\left( {\frac{l}{{{l_{Pl}}}}} \right)^{1/3}},\quad {l_{Pl}} \equiv \sqrt {\frac{\hbar }{\eta }} c
\end{equation}
Similar relations can be obtained for measuring time intervals \cite{s19,s20},
\begin{equation}\label{e19}
  {\left( {\delta t} \right)^2} \ge \frac{{\hbar t}}{{m{c^2}}},\quad \delta t \ge \frac{{Gm}}{{{c^3}}}
\end{equation}
where  $t$  is the measured time interval. By combining these two expressions we find
\begin{equation}\label{e20}
  \delta t \ge {\left( {tt_{Pl}^2} \right)^{1/3}}
\end{equation}
Relation (\ref{e20}) connects the minimum uncertainty during  measurement of time with the measured time interval. The absolute  value of uncertainty   $\delta t \sim {t^{1/3}}$ rises,  whereas its relative value  $\delta t/t \propto {t^{ - 2/3}}$ diminishes. Now rewrite  relations  (\ref{e18}) and (\ref{e20}) in the form
\[\delta l \ge {\left( {\frac{l}{c}\frac{\hbar }{\eta }} \right)^{1/3}}c;\]
\begin{equation}\label{e21}
  \delta t \ge {\left( {t\frac{\hbar }{\eta }} \right)^{1/3}}
\end{equation}
which clearly shows that the existence condition for the minimum uncertainty during measurement of distance (time) is equivalent to the existence condition for the limit power which, in its turn, is dictated by the existence of the  horizon.

To avoid confusion, we emphasize that in the first and in the second case it is not about the accuracy of the particular design of "ruler" or hours, and the universal limitations on the accuracy of the measurement of length and time, which are based on fundamental physical laws.

\section{Principle of maximum force and holographic principle: two principles or one?}

According to the traditional viewpoint,  a dominating  share of the degrees of freedom for our world belongs to the fields that fill the space. But gradually it has become clear  that such an estimation impedes construction of the quantum field  theory. To avoid the problem of divergence at short distances since our world should  be described on a three-dimensional discrete grating  with a period on the order of the fundamental Planck length. Recently another, more radical theory – the so-called holographic principle \cite{s21,s22} – has become popular. Its title is bound up with optical hologram which has the form of two-dimensional image of a three-dimensional object. This principle is based on two key statements:

1. All the information contained in a space domain  can be  "recorded" (presented) on the boundary of this region, which is called the holographic screen;

2. A theory on the boundary of the considered region of space must contain not more than one degree of freedom per Planck area, or in other words total number of degrees of freedom $N$  satisfies the inequality
\begin{equation}\label{e22}
  N \le \frac{A}{{l_{Pl}^2}} = \frac{{A{c^3}}}{{G\hbar }}
\end{equation}
It means that the information density on the holographic screen is limited by the value  $l_{Pl}^{ - 2} = {10^{69}}bit/{m^2}$.

The relation between the number of the degrees of freedom and the volume of the  "memory" of the system is based on the fundamental statement that information always needs a material  carrier: without matter information does not exist.  In the context of this statement  relation  (\ref{e22}) gives an answer to an exceptionally intricate question concerning the maximum density of information record on a material carrier. Following  Susskind \cite{s22}, assume that we have  a physical object with the surface area  $A$ and the entropy $S$. Let this object collapse into a black hole. Naturally, the aria of the horizon of the formed black hole is less than $A$. However, according to thermodynamics of black holes, its entropy cannot decrease herewith.  Accordingly, we obtain the holographic restriction
\begin{equation}\label{e23}
  S \le \frac{A}{{4l_p^2}}
\end{equation}
Here equality is achieved for the objects capable to independently collapse into a black hole.   The relation between the entropy and information makes it possible to formulate the limit density of information record $I \le \frac{A}{{4l_p^2}}$ in the form of holographic limit. In other words, information capacity rises in proportion to the area, but not to the volume. However, such an approach contradicts our intuitive ideas of the relation between  information capacity and the increase of the volume. Below we will show how to calm down our intuition.

As an elementary information carrier, it is convenient to choose the degree of freedom of the system, irrespective of the concrete structure of the matter. Such a choice will make it possible to establish a  relation between the changes in information and those in entropy  expressed directly in terms of the degrees of freedom of the system.

Philosophy of the holographic principle brings forth holographic dynamics in which all (!) the forces known in  nature are replaced by the so-called entropic forces. The latter  are generated by the changes of the information  $I$ on a holographic screen. Taking into account that  $\Delta I =  - \Delta S$ assume that the entropic forces result from the changes in the entropy  due to the displacement  of matter. In other words, the entropic forces originate in the universal tendency of any macroscopic system to increase the entropy. Holographic dynamics may be constructed  in terms of changes in the entropy, and does not depend on details of microscopic theory.  For instance, there is no fundamental field associated with the entropic force.

Consider a physical system of macroscopic size (e.g. Universe)  different parts of which contain certain information. Let us imagine that we have to concentrate some part of the available information in a finite volume during a maximally short period of time. What fundamental limitations will  arise while solving this problem?

Concentration of information in a certain volume is inevitably bound up with concentration of material carriers in the same volume. The existing physical limitations for  energy concentration  lead to direct prohibition of the processes of   excessive information concentration. Most concisely these two types of limitations  are formulated in the form of the principle of maximum force and the holographic principle. Therefore, it seems natural to investigate possible relations between these principles.

Earlier we have shown (see (\ref{e18})) that the existence condition for the minimum uncertainty at measurements of distance is equivalent to the existence condition for the  limit power which, in its turn, is dictated by the existence of horizon. Now reproduce this condition using the holographic principle \cite{s23}.

Let a  volume  ${l^3}$ be partitioned into minimal constituents permissible from the viewpoint of physical laws (e.g. cubes). It seems natural to impart one degree of freedom to each elementary volume (by analogy with the dimensionless cell of phase volume of quantum system $dp \,dq/{\left( {2\pi \hbar } \right)^{3N}}$). If the minimum uncertainty at measurements of the distance  $l$ is $\delta l$, then the elementary volume component  has the volume ${\left( {\delta l} \right)^3}$, and the number of the  degrees of freedom for the system is ${\left( {l/\delta l} \right)^3}$. According to the holographic principle,
\begin{equation}\label{e24}
  {\left( {l/\delta l} \right)^3} \le \frac{{{l^2}}}{{l_{Pl}^2}}
\end{equation}
that immediately sends us back to (\ref{e18}).

It is significant to note that while deriving \cite{s24} (24) we have found the expression for the minimum uncertainty $\delta l$ using  the holographic principle.  As we have earlier seen, the existence of this fundamental characteristic of space is the direct consequence of the principle of the maximum force, and its value can be obtained without taking into account the holographic principle. Therefore, the statement to the effect that the holographic principle is a consequence of quantum fluctuations of space-time \cite{s24} is of the same reliability.

Equivalence of the holographic and the maximum force principles may be also proved quantitatively. As we have earlier seen, the limit density of information on a holographic screen  is bounded by the value  ${\rho _{\inf ,\max }} = \frac{1}{{l_{Pl}^2}} = {10^{69}}bit/{m^2}$. Using the modified Planck length ${l_{Pl}} = \sqrt {\frac{{\hbar {c^2}}}{\eta }} $ we find that $\eta  \approx 9.07 \times {10^{51}}W$. The value of the limit power obtained with the help of the holographic principle  coincides with $\eta  = \frac{{{c^5}}}{{4G}}$ postulated by the principle of maximum force.

\section{Limit relations in information}

The limit values $\hbar ,c,\eta $ control (restrict, bound) the rates of any physical processes, in particular, the rate of information transmission. Its significance is beyond the scope of purely technological applications. To a great extent, the level of development of human society is defined by the rate of information transmission and processing.

One of the initial estimations of the limit rate of information processing is the so-called  Bremermann's limit \cite{s25} $M{c^2}/\hbar  = \left( {M/kg} \right){10^{50}}$bit per second ($M$ is the mass of processor ). For  $M = 1kg$ the time of execution of one operation $\Delta {t_B} \approx {10^{ - 50}}$ sec. is much less than the Planck time, that gives rise to doubts about the estimation adequacy. A weak point of the said estimation is disregard of the processor size and, consequently, the absence of limitation of the rate of transmission of signal inside a computing device.  If to take into account \cite{s26} both the uncertainty principle $\Delta E\Delta t \ge \hbar $ and the finite rate of signal propagation $\Delta t \ge L/c$ , then $\Delta t > \max \left[ {\hbar /M{c^2},L/c} \right]$. Consequently, if the condition  $L < \hbar /Mc$ is fulfilled we can reproduce the  Bremermann's limit $\Delta {t_B} = \hbar /M{c^2}$. However, it is obvious that in this case we  will  also be confronted with the fact that taking into account gravitation makes independent choice of the size $L$ and the  mass $M$. impossible. The processor size is limited by the condition $L > {r_g} = \frac{{2MG}}{{{c^2}}}$ that prevents the formation of a black hole  (horizon). In view of this restriction the minimum time required for execution of one operation
\begin{equation}\label{e25}
  \Delta {t_{\min }} = {\left( {G\hbar /{c^5}} \right)^{1/2}} \sim {10^{ - 43}}\sec
\end{equation}
is the Planck time; the limit rate of any computing device $\nu $ ,
\begin{equation}\label{e26}
  \nu  = t_{Pl}^{ - 1} = {\left( {{c^5}/G\hbar } \right)^{1/2}} \sim {10^{43}}\;bit/\sec
\end{equation}
In terms of limit power, interpretation of the result  (\ref{e26}) is extremely transparent:
\begin{equation}\label{e27}
  \nu  = t_{Pl}^{ - 1} = {\left( {{c^5}/G\hbar } \right)^{1/2}} = {\left( {\frac{\eta }{\hbar }} \right)^{1/2}}
\end{equation}
The rate of information processing by an arbitrary computing device is bounded by the limit concentration of energy inside the device. The limit power $\eta$ is the quantitative measure of this limitation. At $\eta  \to \infty$ (limitation is absent) a computing device could  work at an arbitrarily high rate.

\section{IR-UV Relation}

For description of physical processes with essentially different characteristic scales (of length, time or energy) the traditional point of view dictates to use different approaches. In fact, this means that  macro- and micro-scales are tacitly  assumed to be independent. At present the situation has cardinally changed. It has turned out that even macro-objects possess quantum features. A classical example of such a symbiosis is investigation of black holes. Finding of relations between different scales facilitates  solution of some fundamental problems that have remained unsolved in the framework of traditional approaches.

The new approach \cite{s27} has been called  UV/IR (ultraviolet- infrared)  relation. The hypothesis is based on the following arguments.

In any efficient quantum field theory defined in a spatial domain with the characteristic size   $l$ and using  UV cutoff $\Lambda$, the entropy $S \propto {\Lambda ^3}{l^3}$. Assume additionally that in the framework of the considered theory there are fulfilled the thermodynamic laws of black holes. In particular, this means that in such a theory the entropy $S$ of any object  must be less than the entropy of a black hole  ${S_{BH}}$ of the same size:
\begin{equation}\label{e28}
  S \le {S_{BH}} \approx {\left( {\frac{l}{{{l_{Pl}}}}} \right)^2}
\end{equation}
From here it follows that
\begin{equation}\label{e29}
  {l^3}{\Lambda ^3} \le {\left( {\frac{l}{{{l_{Pl}}}}} \right)^2}
\end{equation}
It is natural to identify the inverse UV-scale with the minimum uncertainty of length measurement $\delta l = {\Lambda ^{ - 1}}$ . In this case   (\ref{e29}) is  immediately transformed into $\delta l = l_{Pl}^{2/3}{l^{1/3}}$.

From the viewpoint of physics of limit values, the relation between small and large scales can be obtained from a  natural condition: the total energy confined within a domain of the  linear size  $l$ must not exceed the energy of a black hole of the same size, i.e.
\begin{equation}\label{e30}
  {l^3}{\rho _\Lambda } \le {M_{BH}} \sim lm_{Pl}^2
\end{equation}
Here  ${\rho _\Lambda }$ is the energy in the volume  ${l^3}$ . Violation of this inequality will lead to the formation of a black hole with the event horizon   preventing further rise of the energy density.
Thus, the relation $\delta l = l_{Pl}^{2/3}{l^{1/3}}$ or its equivalent  (\ref{e29})  for $\delta l = {\Lambda ^{-1}}$ can be considered  the relation between the infrared $l$ and ultraviolet $\delta l$  scales in effective quantum field theory  in which the thermodynamic laws of black holes are fulfilled.

It is interesting to note that the UV/IR relation has broken down  [see (\ref{e28})] seemingly  unquestionable statement to the effect that the obtaining of  information about the structure of a spatial object of the size $\Delta x$ requires the energy
\begin{equation}\label{e31}
  {E_{\Delta x}} \approx \frac{{\hbar c}}{{\Delta x}}
\end{equation}
Relation (\ref{e31}) dictates an evident strategy:  investigation of smaller spatial scales necessitates construction of more and more powerful accelerators. Such a strategy does not take into account gravitational effects and is not doubt on spatial scales essentially exceeding the Planck ones. What will happen if we achieve energies of the order of the Planck ${E_P} = {m_{Pl}}{c^2}\;$ and higher? Accelerators of such energies will turn out to be pointless. It will be impossible to analyze the result of collision of high-energy particles, since the collision products will be hidden by  horizon of the radius
\begin{equation}\label{e32}
  {R_S} = \frac{{2G}}{{{c^2}}}\frac{{{E_{\Delta x}}}}{{{c^2}}} = \frac{{2G}}{{{c^4}}}{E_{\Delta x}} \approx \frac{{{E_{\Delta x}}}}{{{F_{\max }}}}
\end{equation}
Due to finiteness of ${F_{\max }}$ a giant "Super Plancketron" collider \cite{s28} will not permit to obtain  information  about spatial scales which are less than the Planck scale, no matter how high the energy of this accelerator may be.

\section{Maximal acceleration}

The existence condition for the traditional space-time in the presence of  vacuum polarization (virtual processes of production and annihilation of pairs caused by quantum fluctuations) leads to limitation of proper acceleration relatively to the vacuum, or, in other words, to the occurrence of the maximal acceleration \cite{s29,s30,s31,s32}.

The  proper  acceleration    of  the particle $a$ in  curved  space-time is the scalar defined by the relation
\begin{equation}\label{e33}
  {a^2} =  - {c^4}{g_{\mu \nu }}\frac{{D{v^\mu }}}{{ds}}\frac{{D{v^\nu }}}{{ds}}
\end{equation}
where ${g_{\mu \nu }}$ is the metric  tensor, ${v^\mu } \equiv d{x^\mu }/ds$, the dimensionless four-velocity  of the particle, $D/ds$ is the  covariant  derivative  with respect  to  the line element on the world line of the particle,
\begin{equation}\label{e34}
  \frac{{D{v^\mu }}}{{ds}} \equiv \frac{{d{v^\mu }}}{{ds}} + \Gamma _{\alpha \beta }^\mu {v^\alpha }{v^\beta }
\end{equation}
Here $\Gamma _{\alpha \beta }^\mu$ are the affine connections  (the Christoffel symbols) of  space-time  with the metric  ${g_{\mu \nu }}$ , $d{s^2} = {g_{\mu \nu }}d{x^\mu }d{x^\nu }$, the linear element of this space-time.

From the energy-time uncertainty principle it follows that the lifetime of the virtual pair  particle-antiparticle (with the particle mass $m$) generated due to vacuum fluctuations, is  $ \approx \hbar /2m{c^2}$, whereas the distance covered during this time is $ \approx \hbar /2mc$ (the Compton wavelength of the particles). If a virtual particle acquires  the energy equal to its rest mass, it will be transformed into a real particle. When considering the rest system of a particle which is, generally speaking, non-inertial, we find that it undergoes the  inertial  force  ${F_{in}} = \left| {ma} \right|$, where $a$ is  the proper particle acceleration. The work executed by the inertial force during the particle lifetime  $A = ma \times \frac{\hbar }{{2mc}}$. If  $A = m{c^2}$, then there arises  acceleration
\begin{equation}\label{e35}
  a = \frac{{2m{c^3}}}{\hbar }
\end{equation}
At this acceleration, particles of the mass  $m$ will be copiously produced from  the vacuum.  The growth of acceleration will lead to the rise of the mass of the produced particles. What critical consequences may  arise at unlimited growth of acceleration?  If the value of acceleration is high enough, the produced particles can  be transformed into black holes. This will occur in the case when the Compton wavelength of a particle  (particle "size") $\hbar /mc$ is less than its  Schwarzschild radius $2Gm/{c^2}$,
\begin{equation}\label{e36}
  \hbar /mc < \frac{{2Gm}}{{{c^2}}}
\end{equation}
From here it follows that the threshold for black hole formation is a mass of the order of the Planck mass  ${\left( {\hbar \eta } \right)^{1/2}}/{c^2}$. By substituting  $m = {m_{Pl}}$ into  (\ref{e1}) we find
\begin{equation}\label{e37}
  {a_0} \approx \sqrt {\frac{\eta }{\hbar }\,} c
\end{equation}
(as before, we omit the multipliers of the order of unity). At such an acceleration, production of black holes with the  Planck mass due to vacuum polarization will result in breakdown of the traditional knowledge of the structure of space-time, and the acceleration concept itself will lose its conventional sense.  Therefore, the value  ${a_0}$ should be considered the maximal proper acceleration relatively to the vacuum. Note that the presence of the maximum acceleration leads to the formation of a horizon even in SR. In fact, from the viewpoint of SR, the length $l$ of an object moving with the acceleration  $a$ is limited by the relation  $l \le \frac{{{c^2}}}{{2a}}$. On the other hand, it cannot be less than  $l \ge {l_{Pl}} = \sqrt {\frac{\hbar }{\eta }} c$. When  using this inequality for acceleration one obtains  $a \le c\sqrt {\frac{\eta }{\hbar }}$. As is seen, the maximal acceleration corresponds to the fundamental acceleration in the Planck system of units, and is a simple combination of the three limit values $\hbar, c, \eta $.  The necessary condition for its existence is finiteness of all the three limit values : at $c \to \infty $, $\hbar  \to 0$ or  $\eta  \to \infty $ the maximum acceleration is absent.

The presence of the maximum proper acceleration ${a_0}$ (33) automatically leads to the existence of the  minimum  radius of curvature ${R_{\min }}$ of the particle world lines. The radius of curvature of the world line is $R = {c^2}/a$ (since the centripetal acceleration during motion along  the circle of the radius $R$ is $a = {v^2}/R$). Therefore, the minimum radius of curvature has the form
\begin{equation}\label{e38}
  {R_{\min }} = \frac{{{c^2}}}{{{a_0}}} \approx {\left( {\frac{{\hbar G}}{{{c^3}}}} \right)^{1/2}} = c{\left( {\frac{\hbar }{\eta }} \right)^{1/2}}
\end{equation}
Again, we clearly see the key role of the horizon which produces the limit power and, as a consequence,  the maximal proper acceleration and the minimum radius of curvature of the world line.

\section{An ideal quantum clock and Principle of maximum force}

Achievement of required accuracy in any quantum measurement inevitably imposes certain limitations on characteristics of the device designed to perform it. All possible methods to measure the time always involve observation of some periodical physical process. As an example (following \cite{s33}), consider a quantum clock based on observation of radioactive disintegration described by the following equation
\begin{equation}\label{e39}
  \frac{{dN}}{{dt}} =  - \lambda N
\end{equation}
where $N(t)$ is the current number of radioactive particles in the sample. Average number of the decayed particles  during the time interval $\Delta t \ll {\lambda ^{ - 1}}$ is $\Delta N = \lambda N\Delta t$. It enables us to measure the time intervals calculating number of the decaying particles
\begin{equation}\label{e40}
  \Delta t = \frac{{\Delta N}}{{\lambda N}}
\end{equation}
The relative error of such a method of time measurement $\varepsilon  = {\left( {\lambda N\Delta t} \right)^{ - 1/2}} = 1/\sqrt {\Delta N}  \le 1$. At first  glance, it seems that increasing size of the quantum clock (the number $N$), we would gain unlimited improvement in accuracy of the time interval measurement. However, such a process is limited by the following condition: the rise of the clock mass must not lead to transformation of the clock into a black hole (i.e. to occurrence of  horizon). Let us analyze the quantitative limitations  which may be caused by this condition.
By using the uncertainty principle $\Delta E\Delta t \ge \hbar /2$ we can transform  (\ref{e40}) into the inequality
\begin{equation}\label{e41}
  \Delta t \ge \frac{\hbar }{{2{\varepsilon ^2}{c^2}}}\frac{1}{M}
\end{equation}
where $M = N{m_p}$ (with ${m_p}$ corresponding to  the mass of one particle) is the clock mass. If the clock radius $R$ (the clock is assumed to be spherical) becomes less than the gravitational radius  ${R_g}$, it will be impossible to use the clock for time measurements. The condition  $R > {R_g}$ is transformed into
\begin{equation}\label{e42}
  \frac{1}{M} > \frac{{2G}}{{{c^2}R}}
\end{equation}
When substituting  (\ref{e42})  into (\ref{e41}) we obtain
\begin{equation}\label{e43}
  \Delta tR > \frac{1}{{{\varepsilon ^2}}}\frac{G}{{{c^4}}}\hbar
\end{equation}
Treating $R$  as uncertainty $\Delta r$  in position of the physical object (the clock), which is the basis for the time measurement process, and taking into account that $\varepsilon  \le 1$ , one finally obtains \cite{s33}
\begin{equation}\label{e44}
  \Delta t\Delta r > \frac{G}{{{c^4}}}\hbar
\end{equation}
The obtained inequality limits the possibility to determine the time and space coordinates of events to an arbitrary precision.

Let us analyze expression (44) using the notion of the limit force \cite{s34}. For this purpose present it in the form
\begin{equation}\label{e45}
  \Delta t\Delta r > \frac{1}{{{F_{\max }}}}\hbar
\end{equation}
At the fixed Planck constant $\hbar$, it is only the limit force ${F_{\max }}$  defines the limitation  imposed on the quantum clock size. If  such a force is absent in the theory, i.e. ${F_{\max }} = \infty $, then ${R_g} \to 0$, and limitation for the quantum clock size is absent too. The main cause of the discussed limitation is the requirement $R > {R_g}$ equivalent to the condition preventing the formation of  horizon. Therefore, the  occurrence of the force  ${F_{\max }}$ in relation (\ref{e45}) which can be achieved only at the horizon seems absolutely natural.

The structure of relation (\ref{e45}) does not contain any information concerning the process  which has been the base for construction of the clock. This suggests the idea that this relation may be obtained from general considerations. To prove this statement let us use the uncertainty relation
\begin{equation}\label{e46}
  \Delta {x_{\min }}\Delta {p_{\max }} \ge \frac{\hbar }{2}
\end{equation}
Since  ${F_{\max }} = \frac{{\Delta {p_{\max }}}}{{\Delta {t_{\min }}}}$ we immediately obtain that the minimum size $\Delta {x_{\min }}$ of the clock necessary for measurement of the time intervals  $\Delta {t_{\min }}$ obeys the limitation
\begin{equation}\label{e47}
  \Delta {x_{\min }}\Delta {t_{\min }} \ge \frac{\hbar }{{{F_{\max }}}} = \frac{{\hbar c}}{\eta }
\end{equation}
in complete correspondence with (45). This is just the relation that describes the structure of space-time foam! A simple form of relation  (47) points to the fact that  the limit values $\hbar ,c,\eta $ have a fundamental character.

Certainly, the earlier obtained restrictions for the limits of measurability of distance and time  (\ref{e21}) are in accord with relations (47). In fact,  multiplication of the uncertainties  (\ref{e21}) gives
\begin{equation}\label{e48}
  \delta l \cdot \delta t \ge {\left( {\frac{{l\hbar }}{{c\eta }}} \right)^{1/3}}c{\left( {\frac{{t\hbar }}{\eta }} \right)^{1/3}} = {\left( {l \cdot t} \right)^{1/3}}{\left( {\frac{{\hbar c}}{\eta }} \right)^{2/3}}
\end{equation}
Suppose that we are to measure the minimum scales of length and time, i.e. $l = \delta l$ and $t = \delta t$. In such a case (\ref{e48}) will reproduce  relation (\ref{e47}).

\section{Conclusion}

We propose to modify the system of Planck units by replacing the gravitational constant  $G$ by the limit power $\eta  \equiv {c^5}/G$. Such a replacement  makes it possible to use the system of fundamental scales of mass, length and time based only on the limit values  $\hbar, c, \eta $. The latter play a special role in the attempts to describe Nature proceeding from the first principles.

\end{document}